\documentclass[aps,floatfix,eqsecnum,showpacs,preprint,tightenlines]{revtex4}
\usepackage{epsfig}

\usepackage{amsmath}

\usepackage{slashbox}
\usepackage{graphicx}
\usepackage{epstopdf}
\begin{document}

% Use the \preprint command to place your local institutional report
% number in the upper righthand corner of the title page in preprint mode.
% Multiple \preprint commands are allowed.
% Use the 'preprintnumbers' class option to override journal defaults
% to display numbers if necessary
%\preprint{}

%Title of paper
\title{Microscopic Calculation of  $\Lambda-\alpha$ Folding Potential}
\author{Htun Htun Oo}\email{htunhtun.oo93@googlemail.com}
\affiliation{Meiktila University, Meiktila, Myanmar}
\author{Hiroyuki Kamada}\email{kamada@mns.kyutech.ac.jp}
\affiliation{Department of Physics, Faculty of Engineering, Kyushu Institute of Technology, Kitakyushu 804-8550, Japan}

%\author{Evgeny Epelbaum}\email{evgeny.epelbaum@ruhr-uni-bochum.de}
%\affiliation{Institut f\"ur Theoretische Physik II, Ruhr-Universit\"at Bochum, D-44780 Bochum, Germany}

\date{\today}

\begin{abstract}
We construct a folding potential between the $\alpha$ and $\Lambda$  particles based on underlying nucleon-nucleon 
and hyperon-nucleon interactions.
Starting from a phenomenological $\Lambda$-N potential and a Gaussian form of the $\alpha$-particle wave function
we obtain for the built $\alpha$-$\Lambda$ interaction a bound $^5_\Lambda$He state with the binding energy (3.10 MeV), which 
is consistent with recent experimental data $3.12 \pm 0.02$ MeV.
When in turn an exact solution of the four-body Faddeev-Yakubovsky equation for the $\alpha$-particle
calculated with the CDBonn, Nijmegen or Argonne V18 realistic nucleon-nucleon potential is used
and the phenomenological Gaussian $\Lambda$-N potential is replaced by the realistic (Nijmegen NSC97f) potential
approximated by a rank-1 separable form, then $^5_\Lambda$He is overbound. In particular, its binding energy
given by the folding potential generated with the $\alpha$ particle wave function based on the CDBonn potential 
is 7.47 MeV. 
Although the rank-1 separable $\Lambda$-N potential reproduces the exact scattering 
length and the effective range of the original $\Lambda$-N potential,
the overbinding results from the lack of the required repulsive properties
in the assumed separable form.
\end{abstract}

% insert suggested PACS numbers in braces on next line
\pacs{21.45.-v,21.60.Gx,21.80.+a}
%21.60.Gx	Cluster models
%21.80.+a	Hypernuclei
%25.10.+s Nuclear reactions involving few-nucleon systems
%25.40.Cm 	Elastic proton scattering
%21.45.-v 	Few-body systems
%13.75.Cs 	Nucleon-nucleon interactions
% insert suggested keywords - APS authors don't need to do this
%\keywords{Hypernuclei, Folding potential}

%\maketitle must follow title, authors, abstract, \pacs, and \keywords
\maketitle

% body of paper here - Use proper section commands
% References should be done using the \cite, \ref, and \label commands

\section{Introduction}
The strangeness sector of few-body systems still poses many problems despite many years of strong efforts\cite{[1]}.  
Hyperon-nucleon (YN) and hyperon-hyperon (YY) forces are to a large extent unknown due to the sparsity of direct 
YN or indirect YY data.  
Thus the situation is quite different from the nucleon-nucleon (NN) case, where thousands of available data points
strongly constrain the spin-momentum dependencies of nucleon-nucleon force models.  In the case of $\Lambda$-N
or $\Sigma$-N (strangeness S=-1 sector) the set of scattering data is very small \cite{[2]} and is not sufficient to 
determine well the properties of those forces. 
On the theoretical side one is still far away from a solution of the underlying theory of the strong interaction, QCD, 
and therefore effective approaches are used to generate forces.  
They are either based on meson exchanges, like the ones by the Nijmegen \cite{[3]}, 
J\"ulich\cite{ [4]}, or Ehime\cite{ [5]} groups, or on quark models, 
like the ones by the Kyoto-Niigata \cite{[6]}, the Tokyo\cite{ [7]}, the Salamanca \cite{[8]} 
or the Beijing\cite{ [9]} groups.  
Despite all this work, there remains a lot of uncertainty about the properties 
of the baryon-baryon forces, especially in the strangeness S=-2 sector.  

Beside the above-mentioned potentials, there exist also modified versions of the realistic Nijmegen potential, widely used 
in \cite{ [10]}, which are phase equivalent to the original ones and are parameterized in the Gaussian form.  
These YN and YY interactions constitute the dynamical input 
for few-body equations, whose solutions lead to observables, which can be compared to experimental data. 
The lightest hypernuclei $^3_\Lambda$H, $^4_\Lambda$H and $^4_\Lambda$He  with strangeness S=-1 have been 
studied extensively, by evaluating their binding energies and lowest excitation energies by employing various 
potential models.  The results are still ambiguous.  The binding energy of the lightest hypernucleus 
$^3_\Lambda$H can be satisfactorily reproduced using directly some of the Nijmegen forces \cite{ [11]}.  

However, when the Nijmegen YN forces NSC89 \cite{[18]} and NSC97a-f \cite{NSC97f} are employed, no adequate description of $^4_\Lambda$H and 
$^4_\Lambda$He is achieved \cite{[12]}, whereas more phenomenological forces come closer to the data\cite{ [13]}.  
Phenomenological central $\Lambda$-N potentials overbind  $^5_\Lambda$He, which is well known for its 
anomalously small binding energy. This problem seems to be solved by a recent variational 
5-body calculation\cite{[13]} using forces stemming from the original Nijmegen ones.

Certain heavier hypernuclei can also be viewed as few-body systems assuming their cluster structure in terms of 
the $\alpha$-particles.  Phenomenological $\Lambda-\alpha$ potentials of a simple Gaussian type were used e.g. 
to study double-$\Lambda$ hypernuclei \cite{ [14]}.  Recently, we predicted the existence of quasi-bound state of 
the $\Sigma-\Sigma-\alpha$  system employing a phenomenological $\Sigma-\alpha$ potential \cite{[15]}.  
The common drawback of all phenomenological potentials is that they have some parameters to be fit 
to experimental data.  In order to obtain a potential without any unknown parameters, one has to solve a
N-body problem driven by baryon-baryon interactions which is, however, very difficult beyond the four-body system. 
 
In this paper we derive a folding potential between the $\alpha$ and $\Lambda$ particles without 
any additional parameters. Our paper is organized as follows. 
In the next section we explain how to introduce the folding potential in question.
In order to achieve this goal we use the $\alpha$-particle wave function
based on realistic NN forces, e.g., a meson theoretical CD-Bonn \cite{[16]}, 
Nijmegen \cite{Nijm93} and Argonne \cite{AV18} potentials  
and we describe the $\Lambda$-N interaction by a phenomenological Gaussian form \cite{[19]} and 
meson theoretical models, e. g.,
Chiral \cite{chiral}, J\"ulich \cite{Juelich}, Nijmegen \cite{NSC97f,ND} and Ehime\cite{Ehime}. 
In order to facilitate calculations, these meson theoretical $\Lambda$-N potentials 
are modified into convenient separable approximations.
In Sec. \ref{NumericalResult} we calculate binding energies of $^5_\Lambda$He employing
many versions of the folding potential.
Finally, in Sec. \ref{Summary} we discuss and summarize these results.

%%%%%%%%%%%%%%%%%%%%%%%%%%%%%%%%%%%%%%%%%%%%%%%%%%%%%%%%%%%%%%%%%%%%%%%%%%%%%%%%%%%%%%%%%%%%%%%%%%%%%%%%%%%%%%%%%%%%%%%%%%%%%%

%\section{Methodology}
 \section{Method of Calculation}
\label{methodology}
The construction of the $\Lambda-\alpha$ folding potential is performed by using the $\alpha$-particle wave function 
from the solution of the Faddeev-Yakubovsky equations based on realistic NN forces. The $\Lambda$-N interaction 
is first taken in the form of a phenomenological Gaussian potential \cite{[19]}. 
Later we will replace the Gaussian potential by a more realistic one. 
The folding potential $V_{fold}$ is defined by evaluating matrix elements of the inner realistic potential 
$V_{inner}$ between products of two-cluster wave functions $\psi_\alpha \psi_\Lambda$: 
\begin{eqnarray}
V_{fold}=\langle \psi_\alpha \psi_\Lambda | V_{inner} | \psi_\alpha \psi_\Lambda \rangle .
\label{(1)}
\end{eqnarray}
The schematic diagram of the $\Lambda-\alpha$ potential is shown in Fig. 1. For the case at hand, the wave functions $\psi_\alpha$  
and $\psi_\Lambda$ correspond to the $\alpha$ and $\Lambda$ particles.  
The $\alpha$-particle wave function is obtained by solving the Faddeev-Yakubovsky equations \cite{[20]} in momentum space in 
the partial wave basis.  
The natural Jacobi momenta for the $((123)4)\Lambda$ partition \cite{[20]} are 
\begin{eqnarray}
&&
\vec u_1 ={1 \over 2} (\vec k_1-\vec k_2),~~~\vec u_2={2\over 3}\left\{ \vec k_3 -{1\over 2}(\vec k_1+\vec k_2)\right\},
\cr &&
\vec u_3={3\over4}\left\{ \vec k_4 
%\cr &&
-{1\over 3}(\vec k_1+\vec k_2+\vec k_3)\right\},
%\cr &&
\end{eqnarray}
\begin{eqnarray}
%\vec u_\Lambda ={1\over 4m_N+m_\Lambda}\left\{ 4m_N\vec k_\Lambda -m_\Lambda(\vec k_1+\vec k_2+\vec k_3+\vec k_4)\right\},
\vec u_\Lambda ={ 4m_N\vec k_\Lambda -m_\Lambda(\vec k_1+\vec k_2+\vec k_3+\vec k_4) \over 4m_N+m_\Lambda}~~~
\end{eqnarray}
where $\vec k _i, i=1,\dots,4$ are the individual nucleon momenta, $\vec k_\Lambda$ is the momentum of the $\Lambda$ hyperon; $m_N$ 
and $m_\Lambda$ are 
the masses of the nucleon and the $\Lambda$ particle. 
\begin{figure}
\includegraphics[width=.8\textwidth,clip=true]{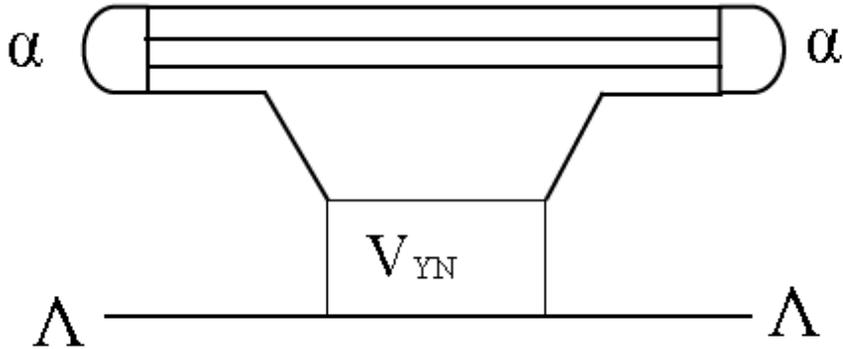}
\caption{A schematic representation of the $\alpha$-$\Lambda$ folding potential.
}
\label{FIG(1)}
\end{figure}
\begin{figure}
\includegraphics[width=.8\textwidth,clip=true]{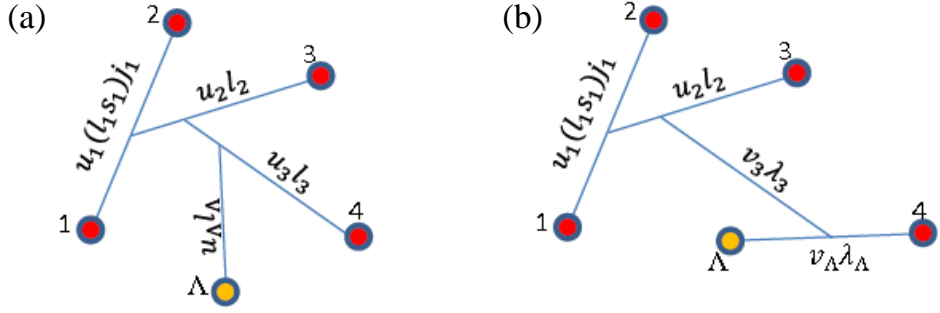} %{image002.eps}
\caption{Definition of continuous and discrete quantum numbers for the $((123)4)\Lambda$ partition 
(left panel) and for the $(123)(4\Lambda)$ partition (right panel).
}
\label{FIG(2)}
\end{figure}
 The corresponding relative orbital angular momenta will be denoted 
by $l_i, i = 1, 2, 3, \Lambda$ and the total spin, total angular momenta and total isospins in the various subsystems by 
$s_i, j_i $ and $ t_i$, 
respectively (see Fig. \ref{FIG(2)}).  The 4N-$\Lambda$ basis states are introduced via
\begin{eqnarray}
&&| u_1 u_2 u_3 u_\Lambda a \rangle  \cr &&
 := | u_1 u_2 u_3 u_\Lambda  \left[l_2 ((l_1 s_1 ) j_1{1\over2}) s_2\right]j_2,  
 (j_2  {1\over 2}) j_3, (l_3 j_3 ) j_\alpha, (l_\Lambda  {1\over 2}) j_\Lambda, 
 (j_\alpha j_\Lambda )J,(t_1  {1\over 2}) t_2 (t_2  {1\over 2})T\rangle ,~~~~
\end{eqnarray}
where the brackets indicate self-explanatory consecutive 
couplings to the total five-baryon angular momentum $J$ 
and total isospin $T$ with the corresponding magnetic quantum numbers (not shown for brevity).  
The quantum numbers for channels $a$ and $b$ are listed in Tab.~\ref{TAB1}. 
The natural Jacobi momenta for the fragmentation $(123)(4 \Lambda)$ are
\begin{eqnarray}
&&\vec v_\Lambda={1\over m_N+m_\Lambda } (m_\Lambda \vec k_4-m_N \vec k_\Lambda ), \cr 
&&\vec v_3={1\over 4m_N+m_\Lambda } 
\{3m_N (\vec k_4+\vec k_\Lambda )-(m_\Lambda+m_N )(\vec k_1+\vec k_2+\vec k_3 )\}.
\end{eqnarray}
The corresponding discrete quantum numbers will be denoted by Greek letters, see the right panel of
Fig. 2. The basis states are 
\begin{eqnarray}
&&| u_1 u_2 v_3 v_\Lambda  b \rangle \cr
&&:=| u_1 u_2 v_3 v_\Lambda  \left[l_2 ((l_1 s_1 ) j_1  {1\over 2}) s_2 \right] j_2,
(\lambda_\Lambda \Sigma_\Lambda ) \tau_\Lambda, (\lambda_3 j_2 ) \tau_3,
(\tau_\Lambda \tau_3 ) J, (t_1  {1\over 2}) t_2 (t_2  {1\over 2})T\rangle ,~
\end{eqnarray}
where the brackets indicate again the sequences of couplings of angular momenta and isospins.  
The Jacobi momenta in these two sets are related via
\begin{eqnarray}
\vec u_3=\vec v_3+{3\over 4} \vec u_\Lambda,~~~~\vec v_\Lambda=-\vec u_\Lambda -{ m_\Lambda \over m_N+m_\Lambda } \vec v_3 .
\end{eqnarray}
In order to calculate the folding potential of Eq. (\ref{(1)}) 
we first prepare the $\alpha$-particle wave function $\psi_\alpha$ and the YN interaction $V_{YN}$ as
\begin{eqnarray}
&&\psi_\alpha (u_1,u_2,u_3, a)=\langle u_1 u_2 u_3 a|\psi_\alpha \rangle, \cr
&& V_{YN} (v_\Lambda,v'_\Lambda )=\langle v_\Lambda |V_{YN}|v'_\Lambda \rangle.
\end{eqnarray}
Then Eq.(\ref{(1)}) turns into
\begin{eqnarray}
&&V_{fold} (u_\Lambda,u'_\Lambda)=4\sum _{a,a'}\int d\vec u_1 d\vec u_2 d\vec u_3 d\vec u_\Lambda \int d\vec u'_1 d\vec u'_2 d\vec u'_3 
d\vec u'_\Lambda \psi_\alpha(u_1 u_2 u_3 a) \psi_\Lambda(u_\Lambda) \cr 
&&\times \langle u_1 u_2 u_3 u_\Lambda  a |V_{YN}|u_1 u'_2 u'_3 u'_\Lambda a'\rangle  \psi_\alpha(u'_1 u'_2 u'_3 a') \psi_\Lambda(u'_\Lambda) \cr
&&=4\sum_{aa'bb'}\int _0 ^\infty v_3^2 dv_3 \int _{-1}^1 dx \int_{-1}^1 dx'  {K_\alpha (u_3 u'_3 a a') \over u_3^{l_3}{  u'}_3^{l'_3} }
 G_{a,b}(u_\Lambda, v_3,x){ V_{YN}(v_\Lambda, v'_\Lambda ) \over {v_\Lambda^{\lambda_\Lambda}{ v'_\Lambda}^{\lambda'_\Lambda} }}
 G_{b',a'} (v_3,u'_\Lambda,x') ,~~~~~~~
 \label{V_fold}
\end{eqnarray}
where 
\begin{eqnarray}
K_\alpha (u_3,u'_3,a,a')=\int d\vec u_1 d\vec u_2 \psi_\alpha(u_1 u_2 u_3 a) \psi_\alpha(u_1 u_2 u'_3 a')
\label{Kernel}
\end{eqnarray}
and $\psi_\Lambda$ is taken is as a plane wave.
The geometrical functions $G_{a,b} (u_\Lambda,v_3,x)$ and $G_{b',a'} (v_3,u'_\Lambda,x')$ 
have been introduced in Refs.~\cite{GloeckleTEXT,Gloeckle1996} 
and for the sake of the reader are displayed in Appendix A.

\begin{table}
\caption{Partial-wave quantum numbers for channels $a$ and $b$ corresponding to Fig.~2.\label{TAB1}}
\begin{tabular}{cccccccccccc}

\hline \hline \noalign{\smallskip}

$a$ & $l_1$ & $s_1$ & $j_1$ & $s_2$ & $l_2$ &  $j_2$  & $l_3$ & $j_3$ & $j_\alpha$ & $l_\Lambda$ & $j_\Lambda$ \\

\hline \noalign{\smallskip} %\hline\noalign{\smallskip}
1 &  0  &  0  &  0  & 1/2 &  0 & 1/2 & 0 & 0   & 0   &  0  & 1/2 \\
2 &  0  &  1  &  1  & 1/2 &  0 & 1/2 & 0 & 0   & 0   &  0  & 1/2 \\
3 &  2  &  1  &  1  & 1/2 &  0 & 1/2 & 0 & 0   & 0   &  0  & 1/2 \\
4 &  0  &  1  &  1  & 3/2 &  2 & 1/2 & 0 & 0   & 0   &  0  & 1/2 \\
5 &  2  &  1  &  1  & 3/2 &  2 & 1/2 & 0 & 0   & 0   &  0  & 1/2 \\
\hline \hline \noalign{\smallskip} 

$b$ & $l_1$ & $s_1$ & $j_1$ & $s_2$ & $l_2$ &  $j_2$  & $\lambda_\Lambda$ & $\sigma_\Lambda$ & $\tau_\Lambda$ & $\lambda_3$ & $\tau_3$ \\

\hline \noalign{\smallskip} %\hline\noalign{\smallskip}
1 &0 &0 &0 &1/2 &0 &1/2 &0 &0 &0 &0 &1/2 \\
2 &0 &1 &1 &1/2 &0 &1/2 &0 &0 &0 &0 &1/2 \\
3 &2 &1 &1 &1/2 &0 &1/2 &0 &0 &0 &0 &1/2 \\
4 &0 &1 &1 &3/2 &2 &1/2 &0 &0 &0 &0 &1/2 \\
5 &2 &1 &1 &3/2 &2 &1/2 &0 &0 &0 &0 &1/2 \\
6 &0 &0 &0 &1/2 &0 &1/2 &0 &1 &1 &0 &1/2 \\
7 &0 &1 &1 &1/2 &0 &1/2 &0 &1 &1 &0 &1/2 \\
8 &2 &1 &1 &1/2 &0 &1/2 &0 &1 &1 &0 &1/2 \\
9 &0 &1 &1 &3/2 &2 &1/2 &0 &1 &1 &0 &1/2 \\
10 &2 &1 &1 &3/2 &2 &1/2 &0 &1 &1 &0 &1/2 \\
11 &0 &0 &0 &1/2 &0 &1/2 &2 &1 &1 &0 &1/2 \\
12 &0 &1 &1 &1/2 &0 &1/2 &2 &1 &1 &0 &1/2 \\
13 &2 &1 &1 &1/2 &0 &1/2 &2 &1 &1 &0 &1/2 \\
14 &0 &1 &1 &3/2 &2 &1/2 &2 &1 &1 &0 &1/2 \\
15 &2 &1 &1 &3/2 &2 &1/2 &2 &1 &1 &0 &1/2 \\
\hline \hline \noalign{\smallskip}
\end{tabular}
\end{table} 

%%%%%%%%%%%%%%%%%%%%%%%%%%%%%%%%%%%%%%%%%%%%%%%%%%%%%%%%%%%%%%%%%%%%%%%%%%%%%%%%%%%%%%%%%%%%%%%%%%%%%%%%%%%%%%%%%%%%%%%%%%%%%%

\section{Numerical results}
\label{NumericalResult}
\begin{figure}
\includegraphics[width=.8\textwidth,clip=true] {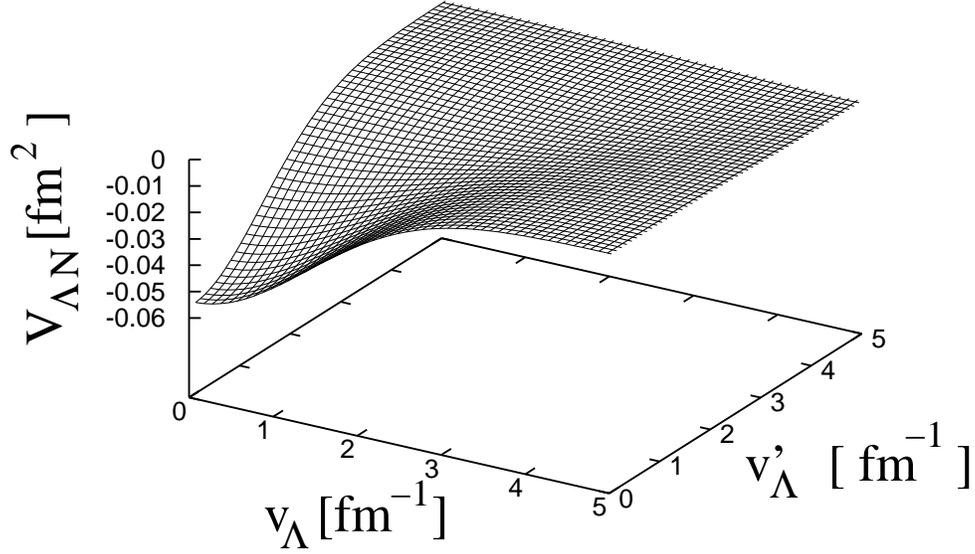} %{fig4.eps}  % {image004.eps}
\caption{Spin triplet $\Lambda$-nucleon interaction in momentum space
}
\label{FIG(4)}
\end{figure}

First let us consider 
results for the $\Lambda-\alpha$ potential obtained with some phenomenological $\Lambda$-N force.
There are many 
phenomenological models of the $\Lambda$-N interaction, 
e.g., a hard-core one represented by exponential functions \cite{[21],[22],[23]} 
or a multi-Gaussian type \cite{[24],Hiyama1997} inspired by the YNG Potential \cite{[25]}:
\begin{eqnarray}
V_{\Lambda N}(r)=\displaystyle \sum_{i=1}^{3}w_{i} e^{-(r/\beta_{i})^{2} } .
\label{(3.1)}
\end{eqnarray}
The spin-dependent phenomenological  $\Lambda$-N interaction is parameterized by Hiyama et al.,\cite{[19]} as
\begin{eqnarray}
V_{\Lambda N}(r)=V_{\Lambda N}^0 (1+\eta~{\bf \sigma}_\Lambda \cdot {\bf \sigma}_N ) e^{-(r/\beta)^2 } ,
\label{(3.2)}
\end{eqnarray}
with $V_{\Lambda N}^0=$-38.19MeV, $\beta$=1.034 fm and $\eta$=-0.1.  
The $\Lambda$-N potential for the spin triplet  is  shown in Fig. \ref{FIG(4)} 
and 
the quantum numbers for channels $a$ and $b$ are listed in Tab.~I  for the total $\mathrm{J}^{\pi}=1/2$  
and the total isospin T=0, assuming the positive parity and restricting to 
$j_{\alpha}$=0, $l_{1}$ = 2, $j_{1}$= 1.  
%Difference between Fig. (3) and (4) is only magnitude. 
The phenomenological potentials of Eqs.~(3.1)--(3.2) are often converted into the $\Lambda-\alpha$ 
potential with the use of 
the Resonating Group Method (RGM) Technique \cite{[14],[26],Hiyama1997}, which is based on the $(0\mathrm{s})^{4}$ 
shell-model Gaussian wave function. Specifically, the integral kernel  $K$  of Eq. (\ref{Kernel}) 
originating 
from the Gaussian $\alpha$ particle S-wave function is given as
\begin{eqnarray}
K(u_3,u_3')=4 \pi ( {2 \over 3 \Omega  \pi }) ^{3\over 2} \exp \{ -{(u_3^2 +u_3'^2) \over 3 \Omega }\} ,
\label{Kernel2}
\end{eqnarray}
where the width parameter $\Omega$ is a common shell model mode \cite{[14],[26]}
taken to be 0.275 fm$^{-2}$.

\begin{figure}
\includegraphics[width=.8\textwidth,clip=true]{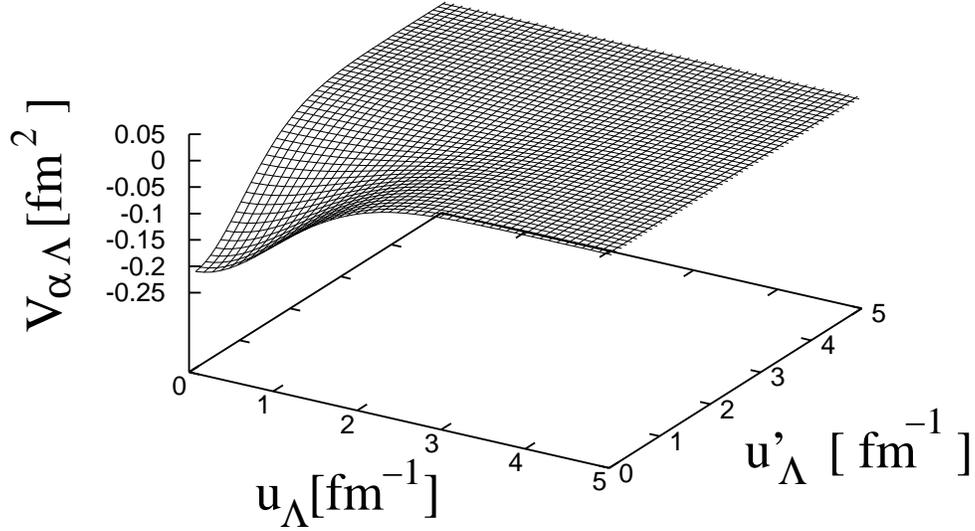} %{fig5.eps}  %{image005.eps}
\caption{$\alpha$-$\Lambda$ folding potential for S-wave.
}
\label{FIG(5)}
\end{figure}

We obtained a $\Lambda$-$\alpha$ folding potential (shown in Fig.\ref{FIG(5)})
for which the ${}^5_\Lambda$He binding energy is -3.10 MeV. 
The calculated  binding energy compares well with the data (-3.12 $\pm$ 0.02 MeV) \cite{Gal1975}.
This consistence is to be expected, since 
the YN potential of Eq. (\ref{(3.2)}) is adjusted to the experimental ${}^5_\Lambda$He binding energy
when the RGM technique is employed.

Next we replace the simple Gaussian 
wave function of the $\alpha$ particle by the wave function based on the realistic NN forces:
the CD-Bonn\cite{[16]}, Nijmegen \cite{Nijm93}, and Argonne V18 \cite{AV18} potentials.  
Our experience with the $\alpha$ particle wave functions obtained with these potentials indicates that 
the S-wave contribution is essential to provide the correct binding energy.
Thus in the following calculations we could restrict ourselves to only few partial waves 
and solve the Faddeev-Yakubovsky equation for the $\alpha$ particle in the basis comprising 
${}^{1}S_{0},\ {}^{3}S_{1}$ and ${}^{3}D_{1}$ states.

On the other hand, the $\Lambda$-N potentials are given in the separable form:
\begin{eqnarray}
V_{\Lambda {\rm N}} (v_\Lambda, v_\Lambda') = -\lambda g(v_\Lambda) g(v_\Lambda') ,
\label{separable}
\end{eqnarray}
where $\lambda$ and $g(p)$ are the coupling constant and the form factor, respectively.
In order to check the accuracy of the separable approximation we prepare two kinds of the separable potentials, e.g, 
the Yamaguchi type (Y) and the Gaussian type (G). The  form factors of these potential are given as  
\begin{eqnarray}
&&g_{\rm Y}(v_\Lambda)= {1 \over v_\Lambda ^2 + \beta _{\rm Y} ^2 },\cr 
&&g_{\rm G}(v_\Lambda)= \exp\{ -  \beta _{\rm G} ^2 v_\Lambda ^2 \}.
\label{separable2}
\end{eqnarray}

The meson theoretical $\Lambda$-N potentials are constructed to describe the $\Lambda$-N scattering data. However,  
due to the sparsity of the data parameters of all potential models are not well determined.
Therefore, we use the fact that the $\Lambda$-N
scattering amplitude in the low energy limit can be determined
from the well-known effective range expansion which has the form
\begin{eqnarray}
k \cot \delta = -{1 \over a} + {1 \over 2} r k^2 + \cdots
\end{eqnarray}
where $k$ denotes the scattering momentum in the center-of-mass system
and
the parameters $a$ [fm]  and  $r$ [fm] are  often called the scattering length and the effective range, respectively.
The phase shift $\delta$ of each partial wave is linked to the scattering amplitude.  

These effective range expansion parameters are directly connected 
to the quantities $\beta _{\rm Y}$, $\lambda_{\rm Y}$,  $\beta _{\rm G}$ and $\lambda_{\rm G}$
from Eq.~(\ref{separable2}) 
by the following relations.
\begin{eqnarray}
&&\beta_{\rm Y} = {3 + \sqrt{9-16{r\over a} } \over 2 r } ,\cr
&& \lambda _{\rm Y} = { 4\beta_{\rm Y}^3 \over \pi \mu (r \beta_{\rm Y}  -1) }, \cr
&&\beta_{\rm G} ={  \sqrt{2}a + \sqrt{a(2a-\pi r) } \over 2\sqrt{\pi}   },  \cr
&&\lambda_{\rm G} = { ar \over {\sqrt{2} \mu \sqrt{a(2a-\pi r)} -(2a-\pi r) \mu } } ,
\label{parameter}
\end{eqnarray}
where $\mu$ is the reduced mass of the $\Lambda$-N system.
In the case of the Yamaguchi type form factors these relations are proved in Ref.~\cite{afnan2015}.
Tables \ref{tab2_} and~\ref{tab2} collect the scattering lengths and the effective ranges 
from several $\Lambda$-N potentials.
Choosing only the $^1S_0$ and $^3S_1$ partial waves of the $\Lambda$N potential 
we can substantially simplify the folding $\Lambda - \alpha$ potential from the 
complicated form given in Eq. (\ref{V_fold}) and arrive at
\begin{eqnarray}
&&V_{fold} (u_\Lambda,u'_\Lambda) \cr
&&=\int _0 ^\infty v_3^2 dv_3 \int _{-1}^1 dx \int_{-1}^1 dx'  K_\alpha (u_3, u'_3)
\{ {1 \over 4} V^{(^1S_0)}_{YN}(v_\Lambda, v'_\Lambda ) + {3\over 4} V^{(^3S_1)}_{YN}(v_\Lambda, v'_\Lambda ) \} .
\label{V_fold_Swave}
\end{eqnarray}
One can note that the spin-spin dependence of the $\sigma_\Lambda \cdot\sigma_N$ Hiyama
$\Lambda$N potential of Eq. (\ref{(3.2)}) 
disappears due to the weighted ($1/4$ and $3/4$) sum in Eq. (\ref{V_fold_Swave}).

\begin{table}
\caption{Scattering lengths $a$ and effective ranges $r$ in fm for $\Lambda$-neutron potential. 
The Hiyama, Chiral and J\"ulich models 
do not differentiate between the $\Lambda$-neutron and the $\Lambda$-proton channel, 
which is indicated with a star($*$).  \label{tab2_}}
\begin{tabular}{ccccc}
\hline \hline \noalign{\smallskip}
Model $\Lambda$N potential                                &  a ($^1S_0$)    &   r  ($^1S_0$)    &  a ($^3S_1$)    &   r  ($^3S_1$) \\
\hline \noalign{\smallskip} 
Hiyama {\it et al.} \cite{[19]}                   &    -1.28*     &  2.33*          & -0.67*       &   3.08* \\
Chiral ($\Lambda$ = 600) \cite{chiral,afnan2015}  &    -2.91*     &  2.78*          & -1.54*       &   2.74* \\
J\"ulich04 \cite{Juelich,afnan2015}               &    -2.56*     &  2.75*          & -1.66*       &   2.93* \\
Nimegen  ESC16                                    &    -1.96      &  3.65           & -1.84        &   3.33 \\
Nijmegen NSC97e \cite{NSC97f,afnan2015}           &    -2.24      &  3.24           & -1.83        &   3.14 \\
Nijmegen NSC97f \cite{NSC97f,afnan2015}           &    -2.68      &  3.07           & -1.67        &  3.34 \\
Nijmegen NSC89 \cite{[18]}                        &    -2.86      &  2.91           & -1.24        &  3.33 \\
 Nimegen  HC-D model \cite{ND,afnan2015}          &    -2.03      &  3.66           & -1.84        &  3.32 \\       
Ehime set 2   \cite{Ehime}                        &     -2.65*    &   3.24*         & -1.80*       &  3.71* \\
Ehime set A   \cite{Ehime}                        &     -2.76*    &   3.19*         & -2.064*      &  3.46* \\
Ehime set B   \cite{Ehime}                        &     -2.71*    &   3.21*         & -1.95*       &  3.56* \\
% \hline  
\hline \hline \noalign{\smallskip}
\end{tabular}
\end{table}

\begin{table}
\caption{Scattering lengths $a$ and effective ranges $r$ in fm for $\Lambda$-proton potential.
\label{tab2}}
\begin{tabular}{ccccc}
\hline \hline \noalign{\smallskip}
Model  $\Lambda$N potential                                &  a ($^1S_0$)    &   r  ($^1S_0$)    &  a ($^3S_1$)    &   r  ($^3S_1$) \\
\hline \noalign{\smallskip} 
Nimegen  ESC16                                    &    -1.88         &  3.58            & -1.86        &   3.37 \\
Nijmegen NSC97e \cite{NSC97f,afnan2015}           &    -2.10         &  3.19            & -1.86        &   3.19 \\
Nijmegen NSC97f \cite{NSC97f,afnan2015}           &    -2.51         &  3.03            & -1.75        &   3.32 \\
Nijmegen NSC89 \cite{[18]}                        &    -2.73         &  2.87            & -1.48        &   3.04 \\
 Nimegen  HC-D model \cite{ND,afnan2015}          &    -2.06         &  3.78            & -1.77        &   3.18 \\       
% \hline 
\hline \hline \noalign{\smallskip}
\end{tabular}
\end{table}

\begin{table}
\caption{ The binding energies of ${}^5_\Lambda$He using the model $\alpha\Lambda$ potentials. The notations (Y) and (G) are corresponding to 
the Yamaguchi separable form and Gaussian one, respectively.
 Unit is in MeV.  \label{tab3}}
\begin{tabular}{c|c|ccccc}
\hline \hline \noalign{\smallskip}
\backslashbox{ $\Lambda$N potential}{NN potential for $\alpha$ particle} & 
\begin{tabular}{c} RGM \\ \cite{Hiyama1997}    \end{tabular} &   
 \begin{tabular}{c} CD-Bonn  \\ \cite{[16]}    \end{tabular} & 
 \begin{tabular}{c} Nijm93   \\ \cite{Nijm93}  \end{tabular} & 
 \begin{tabular}{c} Nijm I   \\ \cite{Nijm93}  \end{tabular} & 
 \begin{tabular}{c} Nijm II  \\ \cite{Nijm93}  \end{tabular} & 
 \begin{tabular}{c} AV18     \\ \cite{AV18}    \end{tabular}  \\
\hline \noalign{\smallskip} 
Hiyama {\it et al.} \cite{[19]}  &   -3.10 %\footnote{The spin-spin dependence is swiched off $\eta=0.0$.} 
&  
                                          -2.99 & -2.18 & -2.54 & -1.98  & -1.95 \\
\hline
Hiyama  (Y)                     & -2.46 & -2.36 & -1.62 & -1.94 & -1.44 & -1.42 \\
Hiyama  (G)                     & -3.07 & -2.89 & -2.12 & -2.46 & -1.92 & -1.90 \\
\hline
  Chiral ($\Lambda$ = 600) \cite{chiral,afnan2015}  (Y) 
                                & -8.44 & -8.26 & -6.64 & -7.36 & -6.21 & -6.16 \\
  Chiral ($\Lambda$ = 600) (G)                                         
                                & -9.26 & -8.83 & -7.42 & -8.08 & -7.03 & -6.99 \\ 
  J\"ulich04 \cite{Juelich,afnan2015}  (Y)     
                                & -8.47 & -8.26 & -6.68 &-7.39 & -6.26 & -6.21 \\
  J\"ulich04 (G)                                   
                                &-9.08 & -8.62 & -7.29 & -7.92 & -6.93 & -6.89 \\

   Nimegen  ESC16    (Y)                                  
                                & -7.57 & -7.27 & -5.98 & -6.57 & -5.65 & -5.61 \\
 Nimegen  ESC16    (G)                         
                                & -7.38 & -6.85& -5.93 & -6.39 & -5.68 & -5.67 \\
Nijmegen NSC97e  \cite{NSC97f,afnan2015}   (Y)  
                                & -8.22 & -7.94 & -6.51 & -7.16 & -6.13 & -6.09 \\
 Nijmegen NSC97e   (G)                            
                                & -8.32 & -7.80 & -6.71 & -7.24 & -6.41 & -6.39 \\
Nijmegen NSC97f \cite{NSC97f,afnan2015} (Y)  
                                & -7.98 & -7.70 & -6.32 & -6.95 & -5.95 & -5.91 \\
Nijmegen NSC97f   (G)                              
                                & -8.00 & -7.47 & -6.44 & -6.95 & -6.16 & -6.14 \\
Nijmegen NSC89 \cite{[18]}    (Y)            
                                & -7.12 & -6.88 & -5.54 & -6.15 & -5.19 & -5.15 \\
Nijmegen NSC89    (G)                          
                                & -7.48 & -7.03 & -5.93 & -6.46 & -5.64 & -5.61 \\
 Nimegen  HC-D model \cite{ND,afnan2015} (Y) 
                                & -7.62 & -7.32 & -6.02 & -6.62 & -5.68 & -5.64 \\
 Nimegen  HC-D model    (G)                      
                                & -7.48 & -6.96 & -6.01 & -6.48 & -5.75 & -5.74 \\   
Ehime set 2 \cite{Ehime} (Y) &    -7.65 & -7.32 & -6.08 & -6.66 & -5.76 & -5.73 \\
Ehime set 2 \cite{Ehime} (G) &    -7.22 & -6.66 & -5.83 & -625& -5.60 & -5.60 \\
Ehime set A \cite{Ehime} (Y) &    -8.79 & -8.45 & -7.05 & -7.70 & -6.68 & -6.64 \\
Ehime set A \cite{Ehime} (G) &    -8.45 & -7.85 & -6.89 & -7.37 & -6.62 & -6.61 \\
Ehime set B \cite{Ehime} (Y) &    -8.31 & -7.97 & -6.64 & -7.26 & -6.29 & -6.26 \\
Ehime set B \cite{Ehime} (G) &    -7.93 & -7.35 & -6.44 & -6.90& -6.19 & -6.18 \\

% \hline 
\hline \hline \noalign{\smallskip}
\end{tabular}
\end{table}

In Tab.~\ref{tab3} we demonstrate the calculated binding energies of the ${}^5_\Lambda$He hypernucleus.
Each row of the table is prepared for one $\Lambda$-N potential. 
The row containing results based on the full potential from Hiyama {\it et~al.}\cite{[19]}) 
is separated from the other ones by a line to indicate 
that the predictions in all other rows are obtained with separable
approximations employing the Yamaguchi type (Y) or the Gaussian type (G) form factors.
Columns tell which realistic NN potential is used to calculate 
the $\alpha$ particle wave function, necessary to construct
the integral kernel $K_\alpha$ in Eq.(\ref{Kernel}) or in Eq. (\ref{Kernel2}).
The column denoted as RGM \cite{Hiyama1997} is an exception because here 
the $\alpha$ particle wave function has a simple Gaussian form,
$\ 2^{-3/4}(\pi \Omega )^{-9/4} \exp \{-(u_1^2+(3/4)u_2^2+(2/3)u_3^2)/(2\Omega)\} $ with $\Omega=0.275$fm$^{-2}$.

From the comparison of the binding energies for the Gaussian wave function 
with the full Hiyama Gaussian $\Lambda$N potential (-3.10 MeV), 
the approximate Hiyama potential (type Y) (-2.46 MeV) and the approximate Hiyama potential (type G) (3.07 MeV),
we can estimate the accuracy of the separable approximation. The accuracy is not better than 
approximately 0.7 MeV.   
Using the realistic $\alpha$ particle wave functions we obtain clear underbinding.
It is most evident for the AV18 potential and predictions based on this NN force 
differ from the others by up to 1.2 MeV.

Surprisingly, the most realistic input for calculations, namely
the realistic $\Lambda$-N potential 
and the $\alpha$ particle wave functions generated by the realistic NN interactions,
leads to rather strong overbinding of the ${}^5_\Lambda$He hypernucleus
and moves the predictions away from the data.
These numbers are listed below the second horizontal line in Tab. \ref{tab3}.

%%%%%%%%%%%%%%%%%%%%%%%%%%%%%%%%%%%%%%%%%%%%%%%%%%%%%%%%%%%%%%%%%%%%%%%%%%%%%%%%%%%%%%%%%%%%%%%%%%%%%%%%%%%%%%%%%%%%%%%%%%%%%%

\section{Summary and Outlook}
\label{Summary}

We have prepared many versions of the $\Lambda$-$\alpha$ folding potential 
which required both $\alpha$ particle wave functions and $\Lambda$-N potentials.
To this end we considered not only the simple Gaussian form of the $\alpha$ particle wave function
but also wave functions generated as rigorous solutions of the four-nucleon Faddeev-Yakubovsky equation
with several realistic NN potentials. 
For the $\Lambda$-N potential we have taken the full and approximated phenomenological
Hiyama model \cite{Hiyama1997} but also 
the meson theoretical ones. 
In order to facilitate our calculations
we prepared and utilized simplified separable versions 
of the $\Lambda$-N realistic potentials, taking care 
to realize exactly their crucial features, the scattering length and the effective range.

First we employed the Hiyama phenomenological $\Lambda$-N potential together with 
the Gaussian $\alpha$ particle wave function
and obtained the binding energy of ${}^5_\Lambda$He -3.10 MeV,
which is in agreement with the experimental data (-3.12 $\pm$ 0.02 MeV).

All the $\Lambda$N potentials used in this paper reproduce both the scattering 
length and the effective range in the  $^1S_0$ and $^3S_1$ states. 
These features are shown in Tabs. \ref{tab2_} and \ref{tab2}. 
Because there is no possibility to compare these parameters with the data, these numbers 
are to some extent arbitrary.
Preserving these features rank-1 separable potentials are prepared as given in 
Eqs.~(\ref{separable})--(\ref{parameter}). The separable approximation of the Gaussian type 
reproduces very well the prediction based on the original phenomenological potential,
yielding the binding energy -3.07 MeV, which is very close to the original -3.10 MeV.

By using the separable approximations both of Yamaguchi and Gaussian type, 
we have obtained may further results for the ${}^5_\Lambda$He binding energy,
which are displayed in Tab.~\ref{tab3}.
It has come as a surprise that for the realistic $\Lambda$N potentials
we get clear overbinding and results are quite different from the data.  
The differences range from 2 MeV to 6 MeV. 
One of the reason of the failure in the description of the data may be the fact
that our rank-1 separable approximation is still unsatisfactory, since 
it realizes only the attractive part of the original potential.  
Presumably, a higher-rank approximation is necessary to account also for 
the repulsive properties of the original potential. 
In near future we plan to introduce a high-rank separable form of the realistic 
$\Lambda$-N potential or to use directly the original, unabbreviated force 
in our calculations.

\acknowledgments
One of authors (H. H. O.) would like to thank Prof. Evgeny Epelbaum for having a chance of collaboration 
physics with DAAD of Germany. We would also like to thank Prof. J. Golak (Jagiellonian Uni.) for helping us with fruitful physics discussions during his visit Japan (2020).
This work was supported partially by Grant-in-Aid for Scientific Research (B) No: 16H04377, 
Japan Society for the Promotion of Science (JSPS).  The numerical calculations were performed on the interactive server at RCNP, Osaka University, 
Japan, and partially on the supercomputer cluster of the JSC, J\"ulich, Germany.

\section*{Appendix A}
\label{appA}
The geometrical functions $G_{a,b} (u_\Lambda,v_3,x)$ and $G_{b',a'} (v_3,u'_\Lambda,x')$ 
were introduced in Refs.~\cite{GloeckleTEXT,Gloeckle1996}
and read:
\begin{eqnarray}
&&G_{a,b} (u_\Lambda,v_3,x)=\sum _k P_k (x) \sum_{L_1+L_2=l_3}\sum_{L'_1+L'_2=\lambda_\Lambda}u_\Lambda ^{L_2+L'_2 } v_3^{L_1+L'_1} 
g_{a,b}^{k,L_1,L_2, L'_1,L'_2 } ,\cr
&&G_{b',a'} (v_3,u'_\Lambda,x')=\sum_k P_k (x') \sum_{L_1+L_2=\lambda'_\Lambda}\sum _{L'_1+L'_2=l'_3}v_3^{L_2+L'_2 } {u'}_\Lambda^{L_1+L'_1} 
g_{b',a'}^{k,L_1,L_2, L'_1,L'_2 } ,
\end{eqnarray}
where the purely geometrical quantities $g_{a,b}^{k,L_1,L_2, L'_1,L'_2 }$ 
and $g_{b,a}^{k,L_1,L_2, L'_1,L'_2 }$ are given by
\begin{eqnarray}
&&g_{a,b}^{k,L_1,L_2, L'_1,L'_2 }={1\over2} \delta_{l_1 l'_1 } \delta_{s_1 s'_1 } \delta_{j_1 j'_1} \delta_{l_2 l'_2 } 
\delta_{j_2 j'_2} \sqrt{ \hat{l_3}  \hat{j_3} \hat{j_\alpha}  \hat{j_\Lambda} \hat{\lambda_\Lambda} \hat{\sigma_\Lambda} \hat{\tau_\Lambda} 
\hat{\tau_3} } (-1)^{\lambda_\Lambda+\sigma_\Lambda} \cr
&&\times
\sum_{LS}\hat{L}\hat{S} %(-1)^{j_2+1+S} 
\left\{\begin{array}{l}
j_{2}\ \frac{1}{2}\ j_{3}\\
\frac{1}{2}\ S\ \sigma_{\Lambda}
\end{array}\right\}\left\{\begin{array}{l}
l_{3}\ j_{3}\ j_{\alpha}\\
l_{\Lambda}\ \frac{1}{2}\ j_{\Lambda}\\
\ L\ S\ J
\end{array}\right\}\left\{\begin{array}{l}
\lambda_{\Lambda}\ \sigma_{\Lambda}\ \tau_{\Lambda}\\
\lambda_{3}\ j_{2}\ \tau_{3}\\
\ L\ \ S\ \ J
\end{array}\right\}\hat{k}\left(\frac{3}{4}\right)^{L_{2}}\left(\frac{m_{\Lambda}}{m_{N}+m_{\Lambda}}\right)^{L_{1}'}
\cr
&&\displaystyle \times\sqrt{\frac{\hat{l_{3}}!}{(2L_{1})!(2L_{2})!}}\sqrt{\frac{\hat{\lambda_{\Lambda}}!}{(2L_{1}')!(2L_{2}')!}}\sum_{ff'}\left\{\begin{array}{l}
L_{1}\ L_{2}l_{3}\\
l_{\Lambda}\ L\ f
\end{array}\right\}\left\{\begin{array}{l}
L_{2}'\ L_{1}'\ \lambda_{\Lambda}\\
\lambda_{3}\ L\ \ f'
\end{array}\right\}C(L_{2}l_{\Lambda}f;00)C(L_{1}'\lambda_{3}f';00) 
\cr
&&
\times\left\{\begin{array}{l}
f\ L_{1}\ L\\
f'\ L_{2}'\ k
\end{array}\right\}C(kL_{1}f';00)C(kL_{2}'f;00),
\end{eqnarray}
and
\begin{eqnarray}
&&g_{b,a}^{k,L_1,L_2, L'_1,L'_2 }={1\over2} \delta_{l_1 l'_1 } \delta_{s_1 s'_1 } \delta_{j_1 j'_1} \delta_{l_2 l'_2 } 
\delta_{j_2 j'_2} \sqrt{ \hat{l_3}  \hat{j_3} \hat{j_\alpha}  \hat{j_\Lambda} \hat{\lambda_\Lambda} \hat{\sigma_\Lambda} \hat{\tau_\Lambda} 
\hat{\tau_3} } (-1)^{l_3+\sigma_\Lambda} \cr
&&\times \sum_{LS} \hat{ L} \hat {S } %(-1)^{1+j_2+S}
\left\{\begin{array}{l}
\frac{1}{2}\ \frac{1}{2}\ \sigma_{\Lambda}\\
j_{2}\ S\ \ j_{3}
\end{array}\right\}\left\{\begin{array}{l}
l_{3}\ j_{3}\ j_{\alpha}\\
l_{\Lambda}\ \frac{1}{2}\ j_{\Lambda}\\
\ L\ S\ J
\end{array}\right\}\left\{\begin{array}{l}
\lambda_{\Lambda}\ \Sigma_{\Lambda}\ \tau_{\Lambda}\\
\lambda_{3}\ j_{2}\ \tau_{3}\\
\ L\ \ S\ \ J
\end{array}\right\}\hat{k}\left(\frac{3}{4}\right)^{L_{1}'}\left(\frac{m_{\Lambda}}{m_{N}+m_{\Lambda}}\right)^{L_{2}}
\cr &&
\displaystyle \times\sqrt{\frac{\hat{\lambda_\Lambda}!}{(2L_{1})!(2L_{2})!}}\sqrt{\frac{\hat{l_3}!}{(2L_{1}')!(2L_{2}')!}}\sum_{ff'}\left\{\begin{array}{l}
L_{1}\ L_{2}\ \lambda_{\Lambda}\\
\lambda_3 \ \ L\ \ f
\end{array}\right\}\left\{\begin{array}{l}
L_{1}'\ L_{2}'\ l_{3}\\
l_\Lambda \ L\ \ f'
\end{array}\right\}C(L_{2}\lambda_{3}f;00)C(L_{1}'l_{\Lambda}f';00)
\cr &&
\times\left\{\begin{array}{l}
f\ L_{1}\ L\\
f'\ L_{2}'\ k
\end{array}\right\}C(kL_{1}f';00)C(kL_{2}'f;00)
\end{eqnarray}
with $\hat x = \sqrt{2x+1}$.

\end{document}